\documentclass{article}

\usepackage{arxiv}

\usepackage[utf8]{inputenc} % allow utf-8 input
\usepackage[T1]{fontenc}    % use 8-bit T1 fonts
\usepackage{hyperref}       % hyperlinks
\usepackage{url}            % simple URL typesetting
\usepackage{booktabs}       % professional-quality tables
\usepackage{amsfonts}       % blackboard math symbols
\usepackage{nicefrac}       % compact symbols for 1/2, etc.
\usepackage{microtype}      % microtypography
\usepackage{lipsum}
\usepackage{graphicx}
\graphicspath{ {./images/} }

\usepackage{multirow}

\usepackage{tabularx}
\usepackage{ragged2e}
\usepackage{array}
\usepackage{tikz}
\usetikzlibrary{shapes.geometric, arrows, positioning, calc}
\usepackage{subcaption}

\title{Towards a unified framework for programming paradigms: A systematic review of classification formalisms and methodological foundations (PREPRINT)}

\author{
  Mikel Vandeloise \\
  University of Namur\\
  Faculty of Computer Science\\ 
  Belgium\\
\texttt{mikel.vandeloise@unamur.be}
}

\begin{document}
\maketitle
\begin{abstract}
The contemporary software engineering landscape, characterized by polyglot programming and the rise of multi-paradigm languages, increasingly challenges traditional methods of language classification. These are not merely theoretical challenges, they manifest as well-documented software engineering problems, including interoperability defects, increased maintenance expenses, and security vulnerabilities. This systematic literature review (SLR) addresses the root of this issue by mapping the formal foundations of programming paradigms. Our objective is twofold: (1) to assess the state of the art of classification formalisms and their documented limitations, and (2) to identify the conceptual primitives and mathematical frameworks proposed for a more powerful, reconstructive approach. \\

Following a rigorous, predefined protocol, we conducted two parallel systematic searches, resulting in the selection and synthesis of 74 primary studies. Our findings for the first research question indicate that existing taxonomies face significant limitations, including a lack of conceptual granularity, an inability to handle hybrid languages, and the absence of a unified formal basis. In response to the second research question, our analysis reveals a strong and sustained convergence within the research community toward a compositional reconstruction of paradigms. This approach focuses on identifying a minimal set of orthogonal, atomic primitives and leveraging sophisticated mathematical frameworks, predominantly Category theory, Type theory, and Unifying Theories of Programming (UTP), to formally guarantee their coherence and compositional properties. \\

We conclude that the literature reflects a notable intellectual shift in the field, moving away from the mature field with well-documented limitations of classification towards the promising future of formal, reconstructive frameworks. This review provides a comprehensive map of this evolution and outlines a concrete research agenda towards the unification of these frameworks. 
\end{abstract}

% keywords can be removed

\section{Introduction}

The contemporary software engineering landscape is characterized by a growing heterogeneity, marked by the adoption of polyglot programming and the emergence of intrinsically multi-paradigm languages \cite{hanus2007multi}. Languages such as Scala, Rust, or F\# natively combine features from functional, imperative, and object-oriented paradigms, offering developers enhanced expressiveness. This sophistication, however, highlights the limitations of traditional conceptual frameworks. Historical taxonomies, which sought to classify languages into distinct paradigmatic categories \cite{ambler1992operational}, are proving increasingly inadequate for capturing the hybrid nature of modern tools \cite{krishnamurthi201913}. This ambiguity can translate into increased complexity and interoperability challenges. These are not merely theoretical concerns, large-scale empirical studies have demonstrated that such inter-language dependencies correlate with higher defect rates and security vulnerabilities in industrial systems \cite{grichi2020impact, kochhar2016large}. Compounding this issue, the very notion of a "paradigm" is often considered ill-defined \cite{michaelsonprogramming}. \\

Given this state of affairs, the fundamental question is no longer merely how to classify languages, but how to understand the first principles that govern them \cite{van2009programming}. The scientific literature lacks a consensus on a "periodic table" of programming concepts, a set of minimal and orthogonal elementary building blocks whose combinations could explain the semantics of different paradigms. The absence of such a unified formal framework \cite{anureev2008problem} not only prevents a rigorous analysis of existing languages but also hinders the design of future ones on solid and coherent semantic foundations. This suggests the need to shift from a perspective of \textit{classification} to one of \textit{reconstruction} \cite{clark1994layered, van2003teaching}. \\

This systematic literature review aims to synthesize the research conducted to address this challenge. Our investigation is guided by the following two research questions:

\begin{itemize}
    \item[\textbf{RQ1:}] What are the existing formalisms and taxonomies for classifying programming languages, and what are their limitations in terms of conceptual granularity and independence from specific implementations?
    \item[\textbf{RQ2:}] How can a set of conceptual primitives be defined to enable the minimal and orthogonal reconstruction of programming paradigms, and how can these properties be formally guaranteed within a unified mathematical framework?
\end{itemize}

In answering these questions, this review makes a twofold contribution. First, it provides a comprehensive and critical assessment of classification approaches, consolidating the arguments for moving beyond them. Second, it identifies and analyses the convergence of research toward a reconstructive approach, highlighting the most promising conceptual primitives \cite{van2009programming} and mathematical frameworks, such as Unifying theories of programming \cite{smith2010embedding} and Category theory \cite{diaconescu1994category}, for founding a unified theory of programming paradigms.\\

The remainder of this article is structured as follows. Section 2 presents the background and fundamental concepts. Section 3 details the rigorous methodology of our systematic review. Section 4 presents the results of our analysis in response to each research question. Section 5 discusses the implications of these findings for research and practice. Finally, Section 6 concludes and proposes avenues for future work.

\section{Background}
\label{sec:Backgroung}

To establish a solid foundation for our review, this section delineates the core concepts that underpin our investigation. We first define programming paradigms, then we underscore the essential role of formal methods in their study.

\subsection{Defining programming paradigms}

A programming paradigm can be understood not merely as a style, but as a coherent set of conceptual tools that shape a programmer's approach to problem-solving \cite{van2009programming}. Seminal work by \cite{wegner1990concepts} describes paradigms as distinct "ways of thinking" that provide a framework for structuring a computational solution. Crucially, these paradigms are often distinguished by the fundamental concepts they choose to include, exclude, or combine. The literature traditionally identifies several major paradigms, each centred on a core abstraction:

\begin{itemize}
    \item \textbf{Imperative programming:} Rooted in the von Neumann architecture of the underlying hardware, this paradigm views computation as a sequence of commands that explicitly manipulate a program's mutable state \cite{reynolds1998}. Its core concepts are the assignment statement, variables representing memory locations, and sequential control flow constructs (loops, conditionals). As such, it is fundamentally operational, describing \textit{how} a result is computed step-by-step \cite{ambler1992operational}.

    \item \textbf{Functional programming:} In contrast, this paradigm models computation as the evaluation of mathematical functions, striving to eliminate side effects and mutable state. Its declarative nature focuses on \textit{what} is to be computed \cite{darlington1992design}. The foundational concepts include first-class and higher-order functions, referential transparency, and the use of recursion for iteration. Its formal basis is often the lambda calculus.

    \item \textbf{Object-oriented programming:} This paradigm organizes software around "objects," which are intended to be self-contained units encapsulating both data (attributes) and behaviour (methods). While widely adopted, its precise definition is a subject of debate. \cite{wegner1990concepts} famously characterized it by the formula: \textit{objects + classes + inheritance}. Others, like \cite{cardelli1984semantics}, place a stronger emphasis on the formal semantics of sub-typing and inheritance as the key mechanisms for achieving data abstraction and polymorphism. The goal is to manage complexity through modular, reusable, and extensible components.

    \item \textbf{Logic programming:} Rooted in formal logic, this paradigm expresses computation as a process of logical deduction over a set of facts and rules \cite{kifer1995logical}. The program specifies logical relationships, and computation is performed by posing queries. The system then attempts to find a proof for the query, using core mechanisms such as unification and backtracking search. This approach fundamentally separates the program's logic from its control flow.

    \item \textbf{Concurrent programming:} This paradigm addresses systems composed of multiple, simultaneously executing computational processes that may interact with one another. Rather than a single paradigm, it encompasses various models for managing this interaction, primarily through shared state (requiring locks and synchronization) or explicit message passing between independent processes \cite{van2009programming}. Its focus is on correctly and efficiently managing communication, synchronization, and resource sharing.
\end{itemize}

While these definitions provide a useful conceptual framework, the rise of multi-paradigm languages demonstrates that their boundaries are not rigid \cite{hanus2007multi}. Modern languages often "freely combine features across supposed boundaries" \cite{krishnamurthi201913}, making a simple classification insufficient. This very blurring of lines motivates the central inquiry of this review: to move beyond these high-level labels and investigate the fundamental, atomic concepts from which they are composed.

\subsection{Conceptual primitives: a working definition}

A central concept in the reconstructive approach to programming languages is the notion of the \textbf{conceptual primitive}. For the purpose of this review, we define a primitive as a fundamental, ideally indivisible, unit of computation \cite{van2009programming}. The goal of many researchers in this area is to identify a set of such primitives that is both \textbf{minimal} (containing the fewest possible concepts) and \textbf{orthogonal} (where each primitive controls an independent aspect of computation), allowing complex language features to be composed of these basic building blocks \cite{van2003teaching, scholz1995four}.\\

It is important to note, however, that the literature does not offer a single, universally accepted definition of what formally qualifies a concept as ``primitive''. As we will show in our discussion section~\ref{sec:Discussion}, clarifying the nature of this term is a key finding of our review. This working definition will serve to ground the initial presentation of our results.

\subsection{The role of formal methods}
\label{sec:The Role of Formal Methods}
To reason about programming languages and their underlying paradigms with precision, the field relies on formal methods, mathematically based techniques for the specification, development, and verification of software systems \cite{abrial1984programming, manna2003mathematical}. Formal semantics, a key component, provides an unambiguous mathematical meaning to program constructs, moving beyond the inherent imprecision of natural language descriptions \cite{reynolds1998}. The choice of a semantic style is not neutral, it often aligns naturally with the core principles of a given paradigm.

\begin{itemize}
    \item \textbf{Operational semantics} defines a program's meaning by specifying its execution on an abstract machine. It describes computation as a sequence of state transitions, making it particularly well-suited for modelling the state changes and sequential control flow that are central to the \textbf{imperative paradigm}. Notably, the \textbf{Structured Operational Semantics (SOS)} framework, pioneered by Plotkin \cite{plotkin1981structural}, provides a powerful compositional method for defining operational semantics, serving as a key precursor to more abstract compositional frameworks.

    \item \textbf{Denotational semantics} associates each programming construct with a mathematical object (e.g., a function) in a semantic domain. By abstracting away from the step-by-step execution, it excels at capturing the compositional and stateless nature of the \textbf{functional paradigm}. The lambda calculus, for instance, serves as the denotational foundation for many functional languages \cite{reynolds1998}.

    \item \textbf{Axiomatic semantics} defines the meaning of commands via logical assertions (e.g., pre- and postconditions) about the program state. While used across paradigms, its primary strength lies in proving program correctness, a concern that transcends any single programming style.
\end{itemize}

Beyond these classical approaches, other formalisms are crucial for modelling the specific features that differentiate paradigms:

\begin{itemize}
    \item \textbf{Type systems and type theory} are fundamental for ensuring safety and structuring programs. By classifying values and expressions, type systems enforce invariants and are central to defining the characteristics of paradigms like object-orientation (e.g., through sub-typing and inheritance semantics \cite{cardelli1984semantics}) and generic programming \cite{willcock2004formalization}. More advanced dependent type systems even allow for the integration of proofs directly within programs \cite{cui2005ats}.

    \item \textbf{Category theory} offers a highly abstract language for describing structure and composition. It allows concepts from different paradigms to be modelled and related as universal algebraic constructions, such as monads for computational effects \cite{ohori1990representing} or indexed categories for logic programming doctrines \cite{amato2009algebraic}.
    
    \item \textbf{Unifying Theories of Programming (UTP)} have been proposed to bridge the semantic gaps between paradigms by using a relational calculus. This approach places different computational models into a common predicate-based framework, enabling unified reasoning about their properties \cite{jifeng2000unifying}.

    \item \textbf{Process calculi and rewriting logic} provide the formal tools to model \textbf{concurrency}. Process calculi, such as the \(\pi\)-calculus, focus on communication and interaction between independent processes. Rewriting Logic, as realized in Maude, models concurrent object systems as the concurrent rewriting of terms, elegantly unifying state change and message passing within a logical framework \cite{meseguer1993logical}.
    
\end{itemize}

By providing a rigorous foundation, this diverse toolkit of formal methods is essential for achieving the goals outlined in our research questions: analysing the precise limitations of existing taxonomies (RQ1) and formally guaranteeing properties like non-contradiction and orthogonality in new reconstructive frameworks (RQ2).

\section{Review methodology}

To provide a comprehensive, transparent, and reproducible overview of the state of the art, this study employs the methodology of a Systematic Literature Review (SLR), following the established guidelines for software engineering \cite{kitchenham2004procedures}. Unlike traditional narrative reviews, an SLR utilizes a strict, predefined protocol to minimize bias in the identification, selection, and synthesis of relevant literature. The protocol governing this review is detailed in the following subsections.

\subsection{Review protocol}
The protocol was designed to rigorously address the two research questions (RQs) that guide this survey.

\subsubsection{Search strategy}
To ensure a comprehensive and replicable search, we developed a detailed search query based on a systematic breakdown of our research questions. We conducted two parallel searches because the vocabulary associated with classification (e.g., `taxonomy`, `ontology`) is substantially different from that associated with formal reconstruction (e.g., `category theory`, `calculus`), making a single combined query less effective and more prone to noise. \\

The query was structured into two primary conceptual groups: \textbf{Group A (Classification)} targeting RQ1, and \textbf{Group B (Formalism \& Reconstruction)} targeting RQ2. Each group was built from a set of core terms and their synonyms, derived from the foundational literature. The final search query combined these two groups with a Boolean OR to capture the full scope of relevant studies. The detailed construction of the search strings is presented in Table~\ref{tab:search_strings}.

\begin{table}[h!]
\caption{Detailed search string construction}
\label{tab:search_strings}
\begin{tabular}{@{}llp{0.7\textwidth}@{}}
\toprule
\textbf{Group} & \textbf{Concept} & \textbf{Keywords / Search string component} \\
\midrule
\multicolumn{3}{l}{\textit{\textbf{Group A: classification (for RQ1)}}} \\
& Core subjects & `programming language` OR `programming paradigm` \\
& \textbf{AND} & \\
& Classification acts & `classification` OR `taxonomy` OR `categorization` OR `typology` OR `feature model` OR `ontology` \\
\midrule
\multicolumn{3}{l}{\textit{\textbf{Group B: formalism \& reconstruction (for RQ2)}}} \\
& Core subjects & `programming language` OR `formalism` OR `paradigm` \\
& \textbf{AND} & \\
& Formal models & `semantics` OR `type theory` OR `calculus` OR `category theory` OR `UTP` OR `Unifying Theories of Programming` OR `formal methods` \\
\midrule
\multicolumn{3}{l}{\textbf{Final combined query}} \\
& \multicolumn{2}{l}{(\textit{Query for group A}) \textbf{OR} (\textit{Query for group B})} \\
\bottomrule
\end{tabular}
\end{table}

\subsubsection{Data sources and period}
The search was conducted across several major academic digital libraries to ensure a comprehensive and balanced coverage of the computer science literature. The selected sources, detailed in Table~\ref{tab:data_sources}, provide access to a wide range of high-impact journals, conference proceedings, and dissertations.

\begin{table}[h!]
\caption{Data sources and rationale for selection}
\label{tab:data_sources}
\begin{tabular}{@{}llp{0.615\textwidth}@{}}
\toprule
\textbf{Data source} & \textbf{Type} & \textbf{Rationale for inclusion} \\
\midrule
\textbf{Semantic Scholar} & \textbf{Indexing DB} & \textbf{The primary database for the initial automated search, providing access to a massive corpus of over 126 million academic papers.} \\
ACM Digital Library & Publisher DB & The primary repository for publications from the Association for Computing Machinery, crucial for foundational and recent work in programming languages. \\
IEEE Xplore & Publisher DB & Provides access to the Institute of Electrical and Electronics Engineers' publications, covering key conferences and journals in software engineering. \\
Scopus & Indexing DB & A large-scale, multidisciplinary abstract and citation database used to ensure broad coverage and help identify relevant work from various publishers. \\
SpringerLink & Publisher DB & Includes numerous influential conferences and journals in theoretical computer science and programming. \\
Google Scholar & Search engine & Used specifically for citation chasing (forward and backward snowballing) to identify foundational or recent papers potentially missed by the primary search. \\
\bottomrule
\end{tabular}
\end{table}

The search covered the period from the inception of each database until May 2025. No start date was set for the search period to ensure the inclusion of foundational papers that have defined the field since its inception, which is essential for a comprehensive historical survey.

\subsection{Study selection process}
The selection of primary studies followed a multi-stage filtering process, guided by a strict set of inclusion and exclusion criteria.

\subsubsection{Inclusion and exclusion criteria}
The selection process was guided by the inclusion (IC) and exclusion (EC) criteria detailed in Table~\ref{tab:inclusion_criteria} and Table~\ref{tab:exclusion_criteria}. These criteria were designed to identify primary studies that directly contribute to answering the research questions, while filtering out irrelevant or non-rigorous work.

\begin{table}[h!]
\caption{Inclusion criteria for study selection}
\label{tab:inclusion_criteria}
\begin{tabular}{@{}llp{0.66\textwidth}@{}}
\toprule
\textbf{ID} & \textbf{Criterion} & \textbf{Description} \\
\midrule
IC1 & Relevance to classification & The study analyses, proposes, or critiques a formal taxonomy or classification system for programming languages or paradigms. \\
IC2 & Relevance to primitives & The study identifies, defines, or formalizes fundamental conceptual primitives of programming. \\
IC3 & Formal foundation & The study presents or utilizes a formal mathematical framework (e.g., logic, calculus, type theory, category theory) to reason about language properties. \\
IC4 & Publication type & The study is a peer-reviewed academic work or evaluated publication (journal article, conference paper, technical report, reviewed preprint or doctoral thesis). \\
IC5 & Accessibility & The full text of the paper is accessible and written in English or French, or translatable satisfactorily into English. \\
\bottomrule
\end{tabular}
\end{table}

\begin{table}[h!]
\caption{Exclusion Criteria for study selection}
\label{tab:exclusion_criteria}
\begin{tabular}{@{}llp{0.66\textwidth}@{}}
\toprule
\textbf{ID} & \textbf{Criterion} & \textbf{Description} \\
\midrule
EC1 & Implementation focus & The study focuses exclusively on the implementation details or performance of a specific language, without generalizing to theoretical principles. \\
EC2 & Non-research contribution & The study is an informal opinion piece, a tutorial, or educational material without a clear research contribution. \\
EC3 & Publication venue & The study is a non-peer-reviewed "grey literature" or an evaluated publication. \\
\bottomrule
\end{tabular}
\end{table}

\subsubsection{Selection procedure}
The selection was performed in three stages. First, the search strings were executed, and all resulting titles and abstracts were screened against the criteria. Second, the full texts of the remaining potentially relevant papers were retrieved and read to make a final inclusion decision. Third, backward, and forward snowballing was performed on the selected papers to identify any additional studies missed by the initial database search.

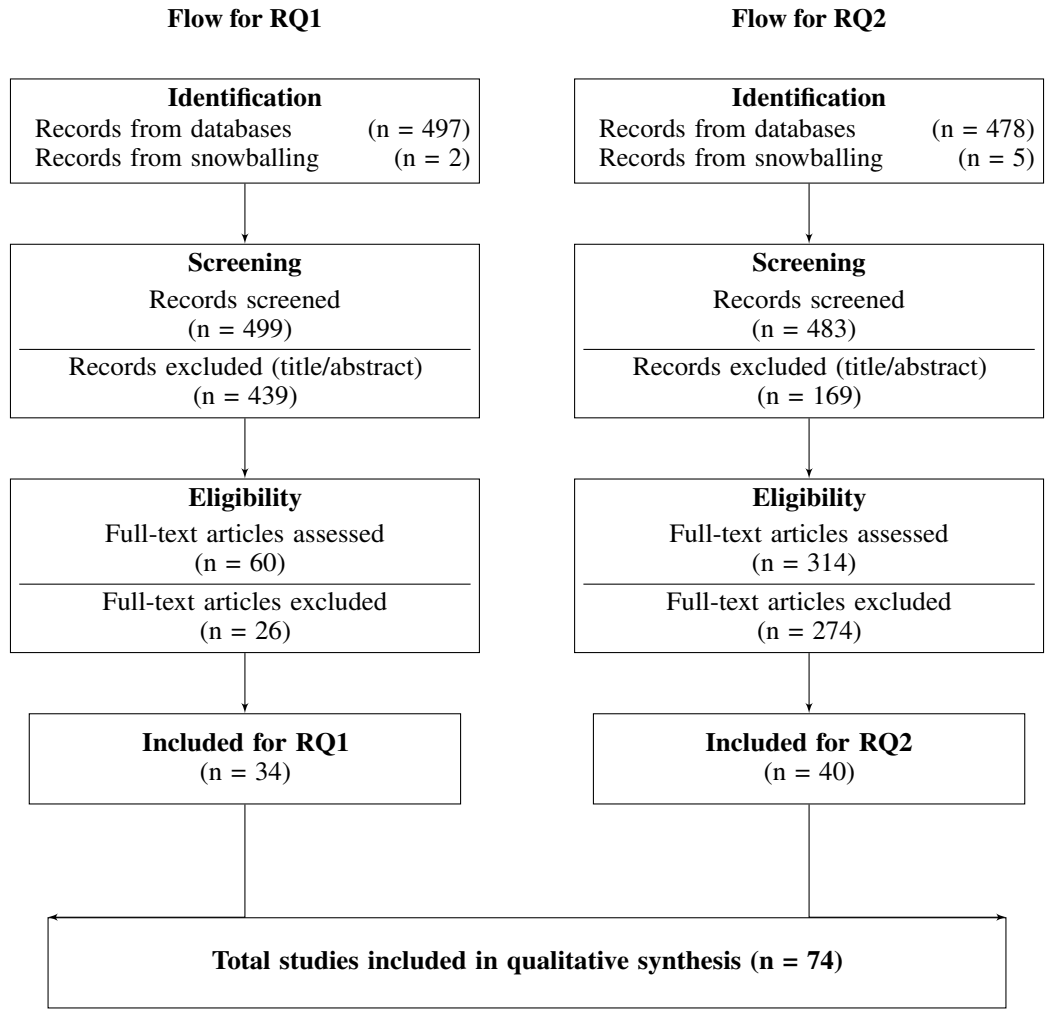
\begin{figure}[h!]
\centering 

\begin{tikzpicture}[
    node distance=0.8cm and 0.5cm,
    box/.style={
        draw,
        rectangle,
        text width=6cm,
        minimum height=1.2cm,
        text centered,
        align=center
    },
    smallbox/.style={
        draw,
        rectangle,
        text width=5.5cm, % Adjusted width
        minimum height=1.2cm,
        text centered,
        align=center
    },
    finalbox/.style={
        draw,
        rectangle,
        text width=12.5cm,
        minimum height=1.2cm,
        text centered,
        align=center
    },
    line/.style={-latex'}
]

% Titles for each flow
\node (title1) at (0,0) {\textbf{Flow for RQ1}};
\node (title2) at (7.5,0) {\textbf{Flow for RQ2}}; % Adjusted position

% --- RQ1 FLOW ---
\node[box, below=0.5cm of title1] (id1) {
    \textbf{Identification} \\
    \vspace{1mm}
    \begin{tabular}{p{4cm} r}
    Records from databases & (n = 497) \\
    Records from snowballing & (n = 2) \\ % Snowballing added
    \end{tabular}
};
\node[box, below=of id1] (screen1) {
    \textbf{Screening} \\
    \vspace{1mm}
    Records screened \\ (n = 499)
    \vspace{1mm}\hrule\vspace{1mm}
    Records excluded (title/abstract) \\ (n = 439) % Adjusted number
};
\node[box, below=of screen1] (elig1) {
    \textbf{Eligibility} \\
    \vspace{1mm}
    Full-text articles assessed \\ (n = 60)
    \vspace{1mm}\hrule\vspace{1mm}
    Full-text articles excluded \\ (n = 26)
};
\node[smallbox, below=of elig1] (inc1) {
    \textbf{Included for RQ1} \\ (n = 34)
};

% --- RQ2 FLOW ---
\node[box, below=0.5cm of title2] (id2) {
    \textbf{Identification} \\
    \vspace{1mm}
    \begin{tabular}{p{4cm} r}
    Records from databases & (n = 478) \\
    Records from snowballing & (n = 5) \\ % Snowballing added
    \end{tabular}
};
\node[box, below=of id2] (screen2) {
    \textbf{Screening} \\
    \vspace{1mm}
    Records screened \\ (n = 483)
    \vspace{1mm}\hrule\vspace{1mm}
    Records excluded (title/abstract) \\ (n = 169) % Adjusted number
};
\node[box, below=of screen2] (elig2) {
    \textbf{Eligibility} \\
    \vspace{1mm}
    Full-text articles assessed \\ (n = 314)
    \vspace{1mm}\hrule\vspace{1mm}
    Full-text articles excluded \\ (n = 274)
};
\node[smallbox, below=of elig2] (inc2) {
    \textbf{Included for RQ2} \\ (n = 40)
};

% --- MERGED TOTAL ---
\node[finalbox, below=1.5cm of $(inc1.south)!0.5!(inc2.south)$] (total) {
    \textbf{Total studies included in qualitative synthesis (n = 74)}
};

% Arrows
\draw[line] (id1) -- (screen1);
\draw[line] (id2) -- (screen2);
\draw[line] (screen1) -- (elig1);
\draw[line] (screen2) -- (elig2);
\draw[line] (elig1) -- (inc1);
\draw[line] (elig2) -- (inc2);

% Merging arrows
\draw[line] (inc1.south) |- (total.north west);
\draw[line] (inc2.south) |- (total.north east);

\end{tikzpicture}%
\caption{PRISMA flow diagram showing the parallel study selection process for RQ1 and RQ2.}
\label{fig:prisma}
\end{figure}

This process resulted in a final set of 74 unique primary studies. The studies are itemized in Appendix A, categorized by the primary research question they address (Table~\ref{tab:rq1_studies} for RQ1 and Table~\ref{tab:rq2_studies} for RQ2). The entire process is summarized in a PRISMA (Preferred Reporting Items for Systematic Reviews and Meta-Analyses) flow diagram (see Figure~\ref{fig:prisma}).

\subsection{Quality assessment}
Each of the finally selected primary studies was assessed for its quality and relevance using the predefined checklist detailed in Table~\ref{tab:quality_criteria}. This step ensured that the subsequent data synthesis was based on methodologically sound and rigorous research. No studies were excluded based solely on this assessment, instead, the results were used to give more weight to higher-quality studies during the data synthesis phase and to contextualize their findings.

\begin{table}[h!]
\caption{Quality assessment criteria}
\label{tab:quality_criteria}
\begin{tabular}{@{}llp{0.68\textwidth}@{}}
\toprule
\textbf{ID} & \textbf{Criterion} & \textbf{Guiding Question} \\
\midrule
QA1 & Clarity of objectives & Are the research objectives or claims of the study clearly stated? \\
QA2 & Context definition & Is the context of the study (e.g., the problem being addressed) clearly defined and motivated? \\
QA3 & Methodological rigour & Is the formal methodology or theoretical framework used in the study rigorously described and justified? \\
QA4 & Strength of evidence & Are the study's findings or contributions well-supported by the presented evidence, arguments, or formal proofs? \\
QA5 & Relevance to RQs & Does the study contribute significantly to the topics addressed by our research questions? \\
\bottomrule
\end{tabular}
\end{table}

\subsection{Data Extraction and synthesis}

A structured data extraction form was developed to collect relevant information systematically from each included study. The form, detailed in Table~\ref{tab:data_extraction}, was designed to capture bibliographic data and specific elements required to answer each research question.

\begin{table}[h!]
\caption{Data extraction form}
\label{tab:data_extraction}
\begin{tabular}{@{}llp{0.72\textwidth}@{}}
\toprule
\textbf{Category} & \textbf{Data item ID} & \textbf{Description and purpose} \\
\midrule
\multicolumn{3}{l}{\textit{\textbf{General information}}} \\
 & DE1 & \textbf{Bibliographic data:} Full citation details (authors, year, title, venue). \\
\midrule
\multicolumn{3}{l}{\textit{\textbf{Data for RQ1: Classification formalisms and limitations}}} \\
 & DE2 & \textbf{Type of formalism/taxonomy:} Description of the proposed classification system (e.g., feature model, ontology, semantic categories). \\
 & DE3 & \textbf{Conceptual granularity:} The level of detail and analytical basis of the classification (e.g., literature-based, theoretical derivation). \\
 & DE4 & \textbf{Identified limitations:} Explicit critiques of existing classification approaches mentioned in the study. \\
\midrule
\multicolumn{3}{l}{\textit{\textbf{Data for RQ2: Reconstruction via primitives + frameworks}}} \\
 & DE5 & \textbf{Conceptual primitives defined:} List of atomic concepts proposed for reconstructing paradigms (e.g., records, named state, concurrency). \\
 & DE6 & \textbf{Mathematical framework:} The formal framework used to guarantee properties (e.g., Category theory, UTP, Rewriting logic). \\
 & DE7 & \textbf{Properties guaranteed:} The specific formal properties ensured by the framework (e.g., orthogonality, non-contradiction, soundness). \\
\bottomrule
\end{tabular}
\end{table}

\subsubsection{Data synthesis}
The extracted data was synthesized using \textbf{thematic synthesis}, a well-established method for analysing qualitative data in systematic reviews \cite{thomas2008methods}. This approach allowed us to systematically identify, analyse, and report patterns (themes) within the data to construct a coherent narrative addressing our research questions. The process consisted of three main phases:

\begin{enumerate}
    \item \textbf{Familiarization and initial coding:} The first phase involved an in-depth reading of the extracted data for all primary studies to achieve a thorough familiarization. During this phase, we performed an initial line-by-line analysis of the extracted text, assigning descriptive codes to key concepts, arguments, and findings relevant to our research questions.

    \item \textbf{Theme development:} In the second phase, we analysed the initial codes to identify similarities and relationships. Related codes were grouped into a hierarchy of descriptive themes. For example, codes such as "ill-defined paradigms", "hybrid language problem", and "lack of objective criteria" were grouped under the higher-level theme "Limitations in Conceptual Granularity". This process was performed iteratively for both RQ1 and RQ2, resulting in two distinct sets of structured themes that accurately reflected the content of the primary studies.

    \item \textbf{Narrative synthesis and interpretation:} The final phase involved generating an analytical narrative from the developed themes. For RQ1, we structured the narrative around the major categories of classification formalisms and the consolidated themes representing their limitations. For RQ2, the themes were used to build a comprehensive map of candidate conceptual primitives and the corresponding mathematical frameworks proposed for their formalization. This structured synthesis enabled us to construct a robust and evidence-based overview of the research landscape and to directly answer our research questions.
\end{enumerate}

\section{Results}
This section presents the results of the systematic review, structured around the two research questions. We first provide a descriptive synthesis of the selected studies (Section~\ref{sec:descriptive_overview}), offering a temporal and venue-based perspective that situates the contributions within their broader research context. We then address RQ1 (Section~\ref{sec:Answer to RQ1}) by analysing the main approaches to programming language classification and identifying their underlying limitations. Finally, we turn to RQ2 (Section~\ref{sec:Answer to RQ2}), examining how various works attempt to reconstruct the space of paradigms from conceptual and mathematical foundations.

\subsection{Descriptive overview of the selected studies}
\label{sec:descriptive_overview}

\begin{figure}[h!]
  \centering
  \includegraphics[width=0.45\textwidth]{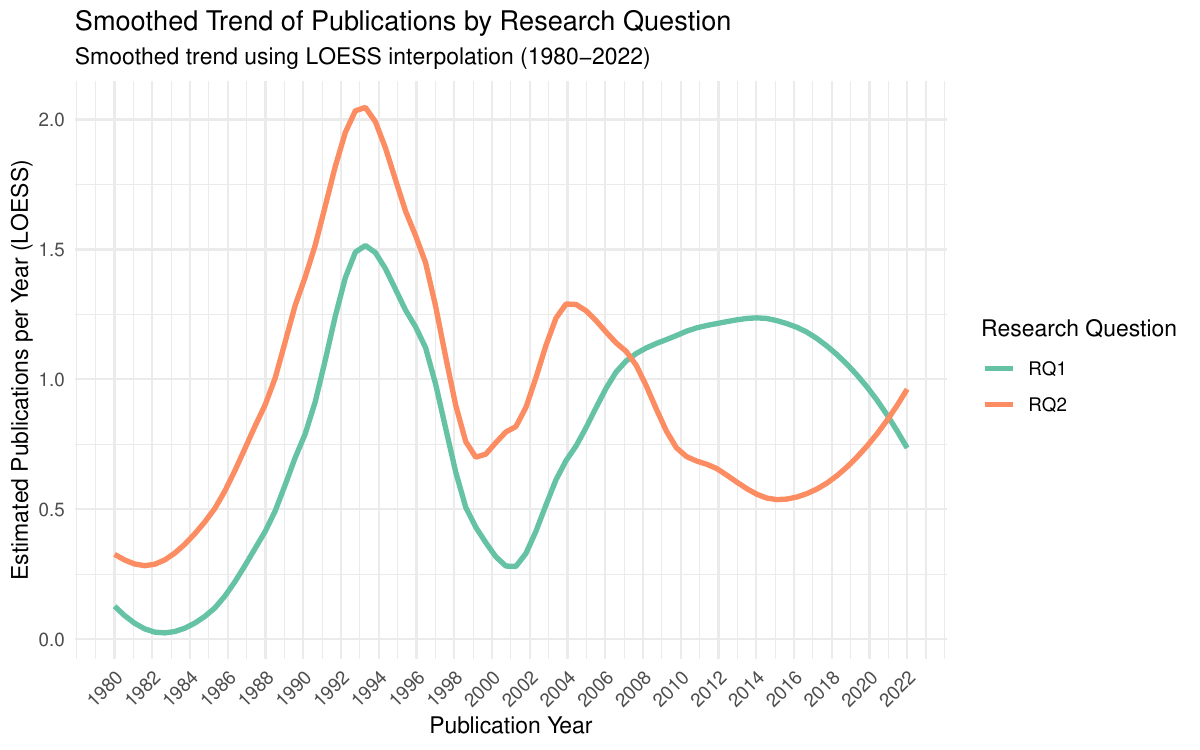}
  \caption{Temporal trends of publications for RQ1 (Classification) and RQ2 (Reconstruction), smoothed using a LOESS model (1980-2022).}
  \label{fig:temporal_trend}
\end{figure}

Before delving into the thematic analysis, this section provides a quantitative overview of the 74 primary studies included in our review. 
Figure~\ref{fig:temporal_trend} illustrates the nuanced dynamics of the intellectual focus over the past four decades, suggesting a cycle of partial substitution between the two research questions. The focus on reconstructive frameworks (RQ2) was dominant throughout the 1990s, peaking sharply around 1994. Following this initial wave, interest in this approach waned, while research on classification and taxonomies (RQ1) gained momentum, becoming the more dominant trend for over a decade, from approximately 2007 to 2020. In the most recent years, the two trends have converged again, with RQ2 regaining significant ground. This cyclical pattern may suggest that the field alternates between periods of foundational, theoretical work (RQ2) and subsequent periods of analysis and consolidation (RQ1). \\

\begin{figure}[h!]
  \centering
  
  \begin{subfigure}[b]{0.45\textwidth}
    \centering
    \includegraphics[width=\textwidth]{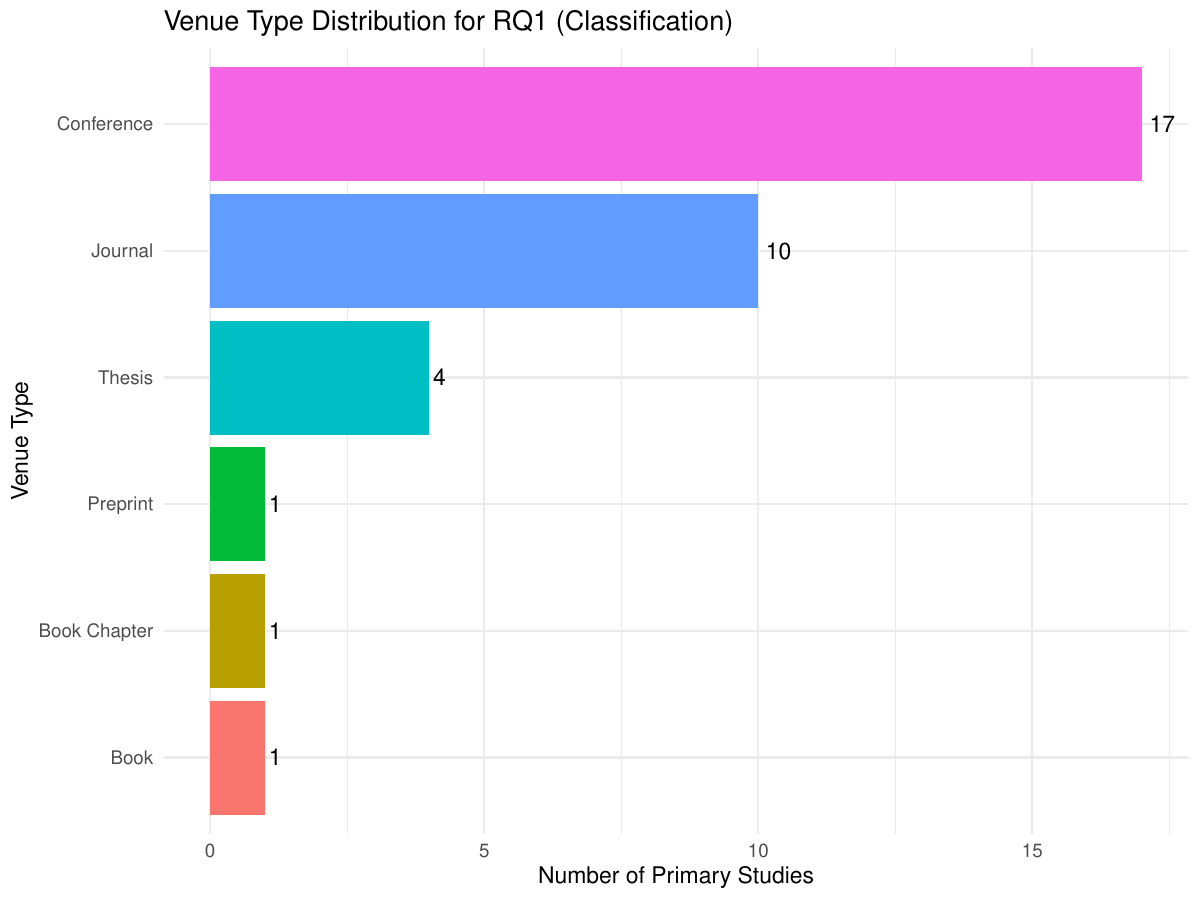}
    \caption{Venue distribution for RQ1 (Classification)}
    \label{fig:venues_rq1}
  \end{subfigure}
  \hfill % Ajoute un espace horizontal flexible entre les deux figures
  \begin{subfigure}[b]{0.45\textwidth}
    \centering
    \includegraphics[width=\textwidth]{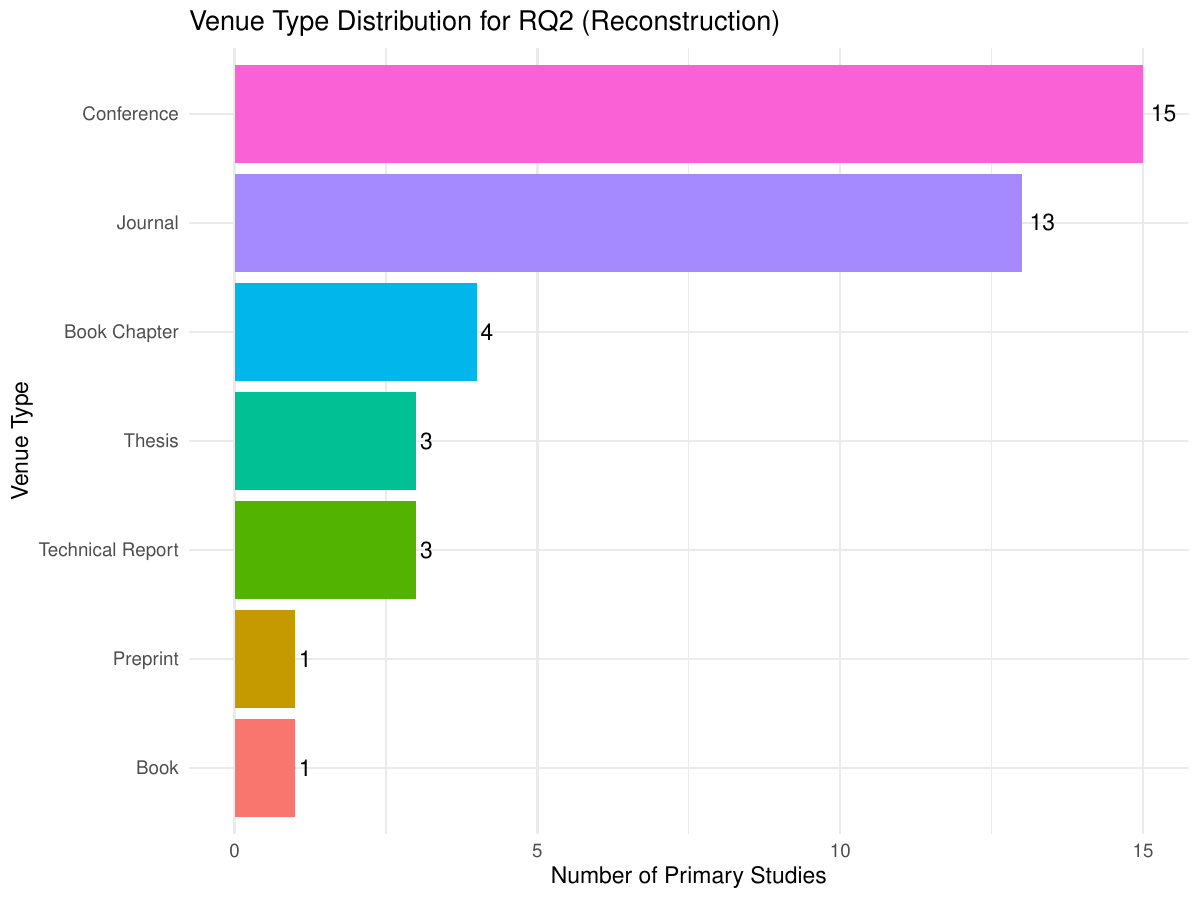}
    \caption{Venue distribution for RQ2 (Reconstruction)}
    \label{fig:venues_rq2}
  \end{subfigure}
  
  \caption{Comparison of publication venue distributions for studies addressing RQ1 and RQ2. Both research streams show a strong presence in conferences and journals.}
  \label{fig:venues_combined}
\end{figure}

The distribution of publication types, as shown in Figure~\ref{fig:venues_rq1} for RQ1 (Classification) and Figure~\ref{fig:venues_rq2} for RQ2 (Reconstruction), reveals broadly similar dissemination patterns. In both cases, conferences and journals constitute the dominant venues, with a slightly higher proportion of conference papers. This consistency suggests that both research strands rely on well-established academic channels for dissemination. However, RQ2 shows a modestly greater diversification across publication types, including book chapters and technical reports, which may reflect a broader engagement with foundational or cross-disciplinary work. Overall, the distribution does not indicate a significant shift in dissemination strategy between the two questions.

\subsection{Answer to RQ1: the state of taxonomies and their limitations}
\label{sec:Answer to RQ1}

Our analysis of the studies addressing RQ1 reveals a strong consensus on the limitations of existing classification methods, despite the variety of formalisms proposed. The findings show a persistent trade-off between conceptual granularity and practical usability \cite{jordan2015feature, krishnamurthi201913}.

\subsubsection{Overview of classification formalisms}
The primary studies propose a range of classification approaches that reveal an evolution from broad, intuitive categories toward more granular and abstract formal models. Our synthesis groups these into four major, often overlapping, categories:

\begin{itemize}
    \item \textbf{Paradigm-based taxonomies:} The most traditional approach, found in numerous foundational studies, uses broad programming paradigms as the primary classification axis \cite{ambler1992operational, steinfeld2014imperative, hanus2007multi}. These taxonomies, which differentiate languages by dominant styles like imperative or functional \cite{anureev2008problem}, offer a high-level, intuitive overview but are consistently criticized for their poorly demarcated boundaries.

    \item \textbf{High-granularity frameworks:} In response to the imprecision of broad paradigms, many researchers developed frameworks with higher conceptual granularity. This includes \textit{feature models} designed to provide objective, fine-grained criteria for comparing language constructs \cite{jordan2015feature}. Similarly, \textit{type-theoretic} and \textit{algebraic models} were proposed to formally capture the specific mechanisms of paradigms, such as inheritance in OOP \cite{lee2015theory, mckenzie1992algebraic}, as well as \textit{object-calculus} approaches that aim to formalize the underlying mechanics of object systems \cite{clark1994layered}.

    \item \textbf{Formal semantic and ontological frameworks:} Seeking maximal implementation independence, a significant body of work relies on highly abstract formalisms. This category includes comprehensive frameworks like the \textit{Unifying Theories of Programming (UTP)}, which use relational calculi to provide a common semantic basis \cite{smith2010embedding}, and \textit{frame logic models} that offer a complete logical foundation for object-oriented systems \cite{kifer1995logical}. More recent approaches leverage \textit{ontology-driven} classifications to create dynamic and evolving formal structures that integrate semantic and pragmatic aspects of languages \cite{anureev2008problem, shilov2011development, akinin2012new}.

    \item \textbf{Empirical and domain-specific frameworks:} A smaller subset of studies develops classification schemes grounded in practice. These frameworks use case studies or empirical data to evaluate and compare languages within specific domains, such as high-performance computing or design applications \cite{el2007productivity, steinfeld2014imperative}, often prioritizing practical applicability over theoretical completeness.
\end{itemize}

Our synthesis reveals clear trends in the characteristics of the reviewed literature. The vast majority of studies employ a formal-theoretical or formal-empirical approach, underscoring the field's commitment to rigour. Furthermore, the body of work is largely balanced between high-granularity models that prioritize formal detail and medium-granularity taxonomies that favour accessibility. While many studies adopt a broad scope covering multiple paradigms, there is a notable concentration of research focused specifically on the object-oriented paradigm. This persistent tension between depth and breadth is a key finding that directly leads to the limitations discussed next.

\subsubsection{Identified limitations}
A central theme emerging from the literature is the inadequacy of existing classification methods to handle the complexity of modern programming languages. The most frequently cited limitations can be grouped as follows:

\begin{itemize}
    \item \textbf{Lack of a unifying formal model:} The most common critique is the absence of a unifying formal model or objective criteria for classification. Researchers note that the term "paradigm" itself is often ill-defined and confusing \cite{krishnamurthi201913, jordan2015feature}, leading to taxonomies with unclear characterizations and differentiations \cite{michaelsonprogramming}. This results in a vague understanding of core concepts like object-orientation, which lacks a single, unifying definition \cite{mckenzie1992algebraic}.

    \item \textbf{Difficulty with hybrid and evolving languages:} The second major limitation is the inability of static classifications to handle the dynamic and hybrid nature of modern languages. Traditional taxonomies struggle with multi-paradigm languages that blend features \cite{hanus2007multi} or with the continuous evolution of languages over time \cite{anureev2008problem}.

    \item \textbf{Trade-off between granularity and usability:} Several works report a fundamental tension: while detailed, high-granularity models gain in formal rigour, they are often complex to apply and may have limited practical applicability \cite{kifer1995logical}. Conversely, broader and more accessible taxonomies necessarily sacrifice nuance and precision \cite{jordan2015feature}.

\end{itemize}

\subsection{Answer to RQ2: foundations for a unified and compositional paradigm theory}
\label{sec:Answer to RQ2}

Taken together, these fundamental limitations highlight the inherent constraints of a purely classificatory approach and help explain the research community's shift towards the reconstructive models based on conceptual primitives that are examined in response to RQ2.

\subsubsection{Candidate conceptual primitives}
A central finding from the studies addressing RQ2 is the convergence toward the \textbf{kernel language approach}, which posits that the vast landscape of programming languages can be understood by focusing on a small, powerful set of fundamental concepts \cite{van2003teaching}. Rather than treating paradigms as monolithic entities, the literature proposes identifying minimal and orthogonal primitives from which more complex features can be reconstructed \cite{clark1994layered, argaranaz1997formalising}.\\

While the exact set of primitives is often tailored to the paradigm being analysed, our synthesis reveals a recurring set of core ideas. Seminal work, for example, identifies concepts such as \textbf{records, lexically scoped closures, independence (concurrency), and named state} as the essential building blocks from which diverse programming styles emerge \cite{van2009programming}. Table~\ref{tab:primitives} provides a synthesized overview of these and other key primitives identified across different paradigm groups in the primary studies. \\

\begin{table}[h!]
\caption{Synthesized conceptual primitives by paradigm group}
\label{tab:primitives}
\begin{tabular}{@{}lp{0.73\textwidth}@{}}
\toprule
\textbf{Paradigm group} & \textbf{Identified core primitives} \\
\midrule
Object-oriented & Object identity, inheritance, encapsulation, dynamic binding \cite{wegner1990concepts, kifer1995logical} \\
Functional & lambda abstraction, recursion, monadic constructs \cite{reynolds1998, ohori1990representing} \\
Logic \& constraint & Set abstraction, proof search, constraints \cite{kifer1995logical, gabbrielli1991modeling} \\
Concurrent \& reactive & Multisets, reaction conditions, locality, channels (often inspired by process calculi) \cite{banuatre2000gamma, oeyen2022graph, scholz1995four} \\
Component \& behavioural & Abstract Behaviour Types (ABTs), data-flow nodes \cite{arbab2005abstract, lepper2014foundations} \\
\bottomrule
\end{tabular}
\end{table}

The overarching goal identified in the literature is not merely to list these concepts, but to define a minimal and orthogonal set that allows for compositional design \cite{van2009programming, scholz1995four}. Notably, while the primitives are diverse, our analysis of the primary studies reveals a significant convergence around the core concepts of both object-orientation (object identity, inheritance, encapsulation) and functional programming (lambda abstraction, monads). This highlights these two paradigms as the primary sources for foundational reconstructive elements and underscores the focus on formal compositionality, which naturally motivates the need for robust mathematical frameworks, as discussed next.

\subsubsection{Mathematical frameworks for formalization}

Our analysis reveals that to formally guarantee properties like orthogonality and correctness, the reviewed studies converge on a sophisticated toolkit of mathematical frameworks. While the landscape of formal methods is broad, as outlined in Section~\ref{sec:The Role of Formal Methods}, our synthesis of the primary studies shows a strong convergence around three particularly prominent families of approaches for the purpose of paradigm reconstruction. It should be noted that formalisms like Process Calculi, while fundamental for concurrency, appeared more frequently in studies dedicated to that specific paradigm, rather than in the core literature focused on a general, unifying framework covering multiple paradigms, which is the focus of RQ2.

\begin{itemize}
    
    \item \textbf{Type theory:} Numerous studies present type theory as essential for ensuring \textit{type safety} and \textit{structural correctness}. The literature provides several key examples of its application, such as providing a formal basis for generic programming concepts \cite{willcock2004formalization}, modelling object hierarchies with extensible tag constructs \cite{lee2015theory}, and enabling the integration of theorem proving directly into programming through dependent types \cite{cui2005ats}.

    \item \textbf{Category theory:} As a frequently cited framework, it is consistently valued for its power to model \textit{compositionality} and \textit{abstraction}. Illustrative applications from our corpus demonstrate its versatility, such as providing a sound semantics for equational and constraint logic programming \cite{diaconescu1994category} and defining algebraic structures for declarative languages using indexed categories \cite{amato2009algebraic}. A central application is the use of monads to represent computational effects like state or concurrency \cite{ohori1990representing}. A major evolution of this monadic approach, \textbf{algebraic effects and handlers}, has since emerged as a more compositional alternative for managing combinations of effects, a theory pioneered by \cite{plotkin2013handling} and further developed in works by \cite{bauer2015programming}.
    
    \item \textbf{Logic-based approaches:} Various forms of logic are used to underpin \textit{algebraic} or \textit{equational reasoning}. This ranges from specific formalisms like Frame Logic (F-Logic), used to prove soundness and completeness in object-oriented systems \cite{kifer1995logical}, to broad relational calculi like the Unifying Theories of Programming (UTP), which is used to model programs as predicates and to mechanize proofs of their properties \cite{santos2006object, nuka2006mechanising}. Unlike the highly abstract and algebraic approach of Category theory which models the structure of composition, UTP adopts a relational and logical approach, focusing on the specification of state-based behaviour via predicates.
\end{itemize}

\subsubsection{A comparative analysis of formal frameworks}
While our results identify the prominent mathematical frameworks used for reconstruction, their respective strengths and trade-offs become clearer when compared to a concrete modelling challenge. To provide this deeper analytical insight, Table~\ref{tab:comparison} illustrates how a single, fundamental concept, \textbf{mutable state}, is modelled within the three major frameworks identified in our review: Type theory, Category theory and Unifying Theories of Programming (UTP). \\

\begin{table*}[h!]
\caption{Comparative modelling of mutable state across formal frameworks}
\label{tab:comparison}
\begin{tabularx}{\textwidth}{@{} l >{\RaggedRight\arraybackslash}X >{\RaggedRight\arraybackslash}X @{}}
\toprule
\textbf{Framework} & \textbf{Core idea} & \textbf{Strengths \& trade-offs} \\
\midrule

\textbf{Type theory} &
State is tracked and controlled directly by the \textit{type system}. This can be achieved via \textbf{effect systems}, where a function's type is annotated with the effects it performs (e.g., \texttt{read(loc)}), or via \textbf{dependent types}, which can model the entire state of the memory as an index to a type. &
\textbf{Strengths:} Provides the strongest static guarantees. Can prevent entire classes of state-related errors at compile-time. Effect systems can offer fine-grained control over which parts of a program are stateful.

\vspace{2mm}
\textbf{Trade-offs:} Can lead to complex and verbose type signatures. May require significant effort from the programmer to satisfy the type checker. \\
\midrule

\textbf{Category theory} &
State is modelled as a computational \textit{effect}. A stateful function $A \rightarrow B$ is transformed into a pure function $A \rightarrow \texttt{State S B}$ that maps an input and an initial state to a result and a final state. Composition is managed algebraically via the \textbf{state monad}. &
\textbf{Strengths:} Highly compositional and abstract. Guarantees that state is properly sequenced. Integrates seamlessly with other effects (I/O, exceptions) if they also have a monadic structure.

\vspace{2mm} % Ajoute un petit espace vertical pour la clarté
\textbf{Trade-offs:} Can become complex when many effects interact (cf. "monad transformers"). The reasoning remains at a high level of abstraction. \\
\midrule

\textbf{UTP} &
State is represented by variables in a formal \textit{alphabet}. A program is a \textbf{relational predicate} that links the values of variables before (e.g., $v$) and after (e.g., $v'$) execution. An assignment $x := E$ is the predicate $x' = E \land y' = y$ for all other variables $y$. &
\textbf{Strengths:} Very expressive and provides a uniform framework for reasoning about imperative programs and their specifications. The logic of pre- and post-conditions is built-in.

\vspace{2mm}
\textbf{Trade-offs:} Can be verbose. The reasoning is very fine-grained and requires managing a potentially large alphabet of variables. \\

\bottomrule
\end{tabularx}
\end{table*}

This comparison reveals that while all three frameworks can formally capture mutable state, they do so at different levels of abstraction and with different goals. Type Theory excels at static verification,  Category theory at algebraic composition and UTP at relational specification. The choice of a "best" framework is therefore not absolute but is contingent on the specific properties a language designer wishes to guarantee, illustrating that the path towards a unified framework may involve integrating concepts from all three approaches.\\

Ultimately, the consensus finding across the reviewed literature is that a carefully selected set of primitives, embedded within one of these unified mathematical frameworks, can achieve modularity, verifiable correctness, and formal guarantees of orthogonality \cite{van2009programming, mycroft1996integration}.

\section{Discussion}
\label{sec:Discussion}

The results of this systematic literature review reveal a significant intellectual shift in the study of programming language foundations. In this section, we synthesize our principal findings, explore their implications for research, practice, and education, and discuss the potential limitations of our study.

\subsection{Synthesis of principal findings}

Our systematic review highlights a significant divergence in the formal study of programming languages, marking a clear intellectual migration from \textit{classification} to \textit{compositional reconstruction}. Two principal findings form the pillars of this discussion. \\

First, our analysis confirms that traditional classification based on monolithic paradigms is a mature field with well-documented limitations. The primary studies provide strong evidence that such taxonomies are often ill-equipped for the modern software landscape. They are consistently criticized for their lack of conceptual granularity, their inability to handle the hybrid nature of multi-paradigm languages, and the absence of a shared formal foundation. This results in a persistent and unresolved trade-off between intuitive simplicity and formal rigour, rendering these models inadequate for deep semantic analysis. \\

Second, and as a direct consequence of these limitations, our findings show a strong and sustained convergence toward a more powerful, reconstructive approach. The literature demonstrates a clear and sustained effort to move beyond mere classification by identifying minimal, orthogonal sets of atomic conceptual primitives. The central thesis of this research front is that paradigms are not indivisible wholes, but rather composites built from these fundamental elements. Our review highlights a clear trend toward leveraging sophisticated mathematical frameworks predominantly Type theory, Category theory and logic-based systems like UTP not just to describe these primitives, but to \textit{formally guarantee} their coherence and compositional properties. \\

As foreshadowed in our background section~\ref{sec:Backgroung}, a key insight from our findings is that the notion of what formally qualifies a concept as "primitive" is not absolute. Our review reveals that the choice of a foundational set of primitives is intrinsically linked to the formal framework used to validate their properties, but rather that the choice of a foundational set of primitives is intrinsically linked to the formal framework used to validate their properties. We can explicitly contrast several distinct "views" on primitivity found in the literature:
\begin{itemize}
    \item The \textbf{programmer-centric view}, emerging from the kernel language approach, where a primitive is a minimal, orthogonal concept that is semantically significant to the programmer, such as `named state` or `lexically scoped closures` \cite{van2009programming}.
    \item The \textbf{algebraic view}, prominent in Category theory, where a primitive is a universal algebraic structure that models a computational feature. A prime example is the `monad`, which serves as the primitive for composing computational effects \cite{ohori1990representing}.
    \item The \textbf{relational view}, central to UTP, where a primitive could be an atomic relational predicate describing a minimal state transition, forming the basis for composing more complex program specifications \cite{jifeng2000unifying}.
\end{itemize}
This suggests that the quest for a "periodic table" of programming is not merely a search for the elements themselves, but also for the formal theory that best endows them with the status of "primitive" by rigorously guaranteeing their properties. This deeper interplay between concepts and their formal validation is a key insight that underpins the entire reconstructive research agenda.

\subsection{Implications for research: A new research agenda}
A key implication of our findings is the crystallization of a paradigm shift in programming language research: a move from a \textit{taxonomic} perspective, classifying existing "species", to a \textit{compositional} one that seeks to map their conceptual "DNA". This shift does not merely open new avenues for research, it proposes a new research agenda. The goal is no longer to fit languages into predefined boxes, but to understand them as engineered artifacts within a formal design space, defined by the principled composition of elementary concepts. \\

Our findings demonstrate that this theoretical pursuit is not an isolated academic exercise. Rather, it is the formal expression of long-standing, practical challenges in software engineering. The difficulty of integrating subsystems written in different paradigms, a problem of "surprising complexity" \cite{mycroft1996integration}, is a direct consequence of the non-orthogonal feature sets of today's languages. This is no longer anecdotal: recent, large-scale empirical studies provide compelling evidence that inter-language dependencies are a significant source of software defects and security vulnerabilities \cite{grichi2020impact, kochhar2016large}. The interoperability failures observed in practice, such as between C and Fortran systems \cite{sultana2016understanding}, are therefore not just bugs, but symptoms of an underlying lack of a sound, unified semantic model. \\

Viewed through this lens, the formal reconstructive frameworks identified in this review represent more than just theoretical models, they are candidate solutions to these deep-seated engineering challenges. Advanced architectural solutions, such as the TruffleVM which enables high-performance interoperability via language-agnostic messages \cite{grimmer2018cross}, can thus be re-interpreted as a practical embodiment of the theoretical quest for orthogonal composition. A key direction for future research, therefore, is to formally connect these two worlds: to use the primitives and frameworks from the theoretical literature to formally model and verify the interoperability mechanisms being developed in practice. \\

\subsection{Implications for practice and education}
The shift towards a reconstructive, compositional model of programming paradigms is not merely a theoretical refinement, it has profound implications for both the practice of language design and the pedagogy of computer science. \\

For language design, our findings suggest a move away from monolithic, feature-driven design towards a more principled, compositional architecture. This approach provides a formal roadmap for creating languages that are not just multi-paradigm, but paradigm-polymorphic, capable of being safely and predictably extended with new computational features. A language could be engineered from a minimal functional core and then systematically extended with formally specified conceptual modules, such as a "concurrency monad" or an "object-capability system", with strong guarantees about how these features will interact. This aligns with and provides a formal basis for research on modular language design, such as structuring paradigms themselves as composable object classes \cite{spinellis1994programming}. \\

This design philosophy has a direct and transformative impact on education. The kernel language approach, identified as a central theme in our results \cite{van2003teaching}, emerges not just as a teaching method, but as the natural pedagogical consequence of compositional design. By focusing on a small set of core, orthogonal concepts first, educators can equip students with a deep, transferable mental model of computation. This method directly addresses the long, standing critique that traditional education often teaches programming as a disparate collection of language, specific crafts \cite{krishnamurthi201913}, instead of fostering a more fundamental and enduring understanding of the principles that unite them.

\subsection{Threats to validity}
To ensure the credibility of our findings, we have critically assessed several potential threats to the validity of this systematic literature review.

\begin{itemize}
    \item \textbf{Construct validity:} A primary threat concerns the construction of our search query. To mitigate the risk of using non-standard or overly restrictive terminology, our search strings were designed to be broad, combining both paradigm-level and formalism-level keywords. The subsequent multi-stage manual screening process, including backward and forward snowballing, further ensured that we captured a comprehensive and representative sample of the relevant literature.

    \item \textbf{Internal validity (researcher bias):} As the data extraction and synthesis were primarily conducted by a single researcher, there is a potential for subjective bias. We implemented two specific measures to counteract this. First, we used a rigidly structured data extraction form with precisely defined fields, minimizing the opportunity for subjective interpretation. Second, the synthesis was performed using a well-established, multi-phase thematic synthesis protocol, which imposes a systematic and traceable process on the analysis of the extracted data.

    \item \textbf{External validity (generalizability):} Our research questions deliberately focus on formal and theoretical studies, meaning our primary corpus does not represent the entirety of programming language literature. This is a necessary limitation to maintain the focus of the review. However, we have explicitly addressed this threat by systematically contextualizing our theoretical findings within the broader, practical landscape of software engineering. As demonstrated throughout the discussion, we have shown that the formal problems identified in our corpus have direct and significant parallels in the empirical literature on real-world challenges like language interoperability and multi-language system maintenance. This strengthens the argument that our findings, while drawn from a theoretical base, are relevant to the wider field.
\end{itemize}

\section{Conclusion and future work}

\subsection{Conclusion}
This systematic literature review was undertaken to map the formal foundations of programming paradigms, motivated by the clear inadequacy of traditional classification schemes. Our findings provide clear answers to our two research questions. \\

First, the literature confirms that existing taxonomies are fundamentally limited by a lack of formal unity and an inability to handle the hybrid, evolving nature of modern languages. Second, in response, our review reveals a strong and sustained convergence within the research community toward a \textbf{reconstructive approach}. This new model is built upon identifying minimal, orthogonal conceptual primitives and leveraging sophisticated mathematical frameworks, notably Type theory, Category theory and UTP, to formally guarantee their compositional properties. \\

Ultimately, this review charts a significant intellectual shift. We conclude that the future of foundational programming language research lies not in refining outdated taxonomies, but in the continued development towards a unified theory based on these powerful, formal, reconstructive frameworks. \\

\subsection{Future work}
The findings of this review illuminate a clear and compelling path for future research. We propose a three-pronged research agenda to build upon the foundations identified in the literature:

\begin{enumerate}
    \item \textbf{Development of a complete unified formal framework:} While our review identified several powerful candidate frameworks (Category theory, UTP, etc.), a comprehensive framework that integrates the full spectrum of promising primitives (e.g., state, concurrency, and logic variables) in a formally verified manner remains an open challenge. Future work should focus on developing such a framework, potentially leveraging the compositional power of category theory and monads to model the interaction of diverse computational effects within a single, coherent mathematical structure.

    \item \textbf{Creation of a reference kernel language:} To validate the expressiveness and utility of such a unified framework, the next logical step is to instantiate it as a concrete, executable reference kernel language. This language would serve as a powerful tool for both research and pedagogy: allowing for the formal specification and verification of new language features, and providing an ideal platform for teaching the fundamental concepts of programming in a unified and language-agnostic manner, as advocated by the kernel language approach.

    \item \textbf{Empirical analysis of multi-language systems:} Finally, the formal models developed must be connected back to the practical challenges of software engineering. We call for new empirical studies that would use the conceptual vocabulary of a unified framework to analyse the complexity and quality of existing multi-language systems. Such studies could, for instance, formally measure the "conceptual friction" at the boundaries between components written in different paradigms, providing a quantitative basis for understanding the software defects and maintenance issues observed in practice.
\end{enumerate}

\section*{Data availability statement}

All data, materials, and analysis scripts used to conduct this systematic literature review are available from the corresponding author upon reasonable request. This includes the full datasets for the screening and extraction phases (CSV files), the R script used to generate the figures presented in this paper, and all supplementary materials. The repository will be made publicly available upon acceptance of the paper.

\bibliographystyle{ACM-Reference-Format}  
\bibliography{references}

\newpage
\appendix
\section{List of Included Primary Studies}

The final set of primary studies selected through our systematic review process is listed below. The studies are grouped according to the primary research question they address. Studies that significantly contribute to both questions may appear in both lists for clarity, though their primary contribution guides their main placement.

\begin{table*}[h!]
\centering
\caption{Primary studies addressing RQ1: taxonomies, formalisms, and their limitations}
\label{tab:rq1_studies}
\scalebox{0.9}{
\begin{tabular}{@{}llp{0.6\textwidth}@{}}
\toprule
\textbf{ID} & \textbf{Reference} & \textbf{Primary focus} \\
\midrule
{[S1.1]} & \cite{akinin2012new} & Ontology-based classification portal \\
{[S1.2]} & \cite{ambler1992operational} & Operational vs. definitional paradigm taxonomy \\
{[S1.3]} & \cite{anureev2008problem} & Semantics and pragmatics view on classification \\
{[S1.4]} & \cite{argaranaz1997formalising} & Formalising sharing mechanisms via object calculus \\
{[S1.5]} & \cite{bono2008traits} & Framework based on separating roles in OOP \\
{[S1.6]} & \cite{breazu1992naturally} & Monad calculus for embedding query languages \\
{[S1.7]} & \cite{cardelli1984semantics} & Formal semantics of multiple inheritance \\
{[S1.8]} & \cite{clark1994layered} & Layered framework for analysing and constructing OOPLs \\
{[S1.9]} & \cite{darlington1992design} & Design space and classification for declarative languages \\
{[S1.10]} & \cite{diaz1996object} & Cross-discipline taxonomy of object-oriented systems \\
{[S1.11]} & \cite{drew1994exception} & Taxonomy of exception handling models \\
{[S1.12]} & \cite{el2007productivity} & Classification of HLLs for HPC perspective \\
{[S1.13]} & \cite{Gorodnyaya2021} & Visual/matrix-based semantic decomposition \\
{[S1.14]} & \cite{gorodnyaya2019method} & Method for paradigmatic analysis by semantic categories \\
{[S1.15]} & \cite{hanus2007multi} & Survey and taxonomy of multi-paradigm declarative languages \\
{[S1.16]} & \cite{jones2017classless} & Formal semantics for classless object-oriented languages \\
{[S1.17]} & \cite{jordan2015feature} & Feature model for comparing language paradigms \\
{[S1.18]} & \cite{kifer1995logical} & F-Logic as a formal foundation for OO/frame-based languages \\
{[S1.19]} & \cite{krishnamurthi201913} & Critique of paradigms; proposal of notional machines \\
{[S1.20]} & \cite{lee2015theory} & Type theory with extensible tags for modelling objects \\
{[S1.21]} & \cite{mckenzie1992algebraic} & Algebraic model for class, message passing, and inheritance \\
{[S1.22]} & \cite{michaelsonprogramming} & Critique of the "paradigm" concept in education \\
{[S1.23]} & \cite{owens2016generic} & A generic framework for cross-paradigm automated quality assurance \\
{[S1.24]} & \cite{paakki1995attribute} & Survey and taxonomy of attribute grammar paradigms \\
{[S1.25]} & \cite{ramos1993design} & Formal semantics (equational logic) for a declarative OOPL \\
{[S1.26]} & \cite{shilov2011development} & Development of an evolving ontology for classification \\
{[S1.27]} & \cite{smith2010embedding} & Critique of paradigms; proposal of Computational Thinking \\
{[S1.28]} & \cite{steinfeld2014imperative} & Programmatic paradigm taxonomy for design applications \\
{[S1.29]} & \cite{teixeira2021multi} & Multi-language static code analysis on the lara framework \\
{[S1.30]} & \cite{valle2013towards} & Semiotic framework for analysing paradigms \\
{[S1.31]} & \cite{van2009programming} & Taxonomy of paradigms based on core concepts \\
{[S1.32]} & \cite{van2002teaching} & The kernel language approach for teaching/unifying paradigms \\
{[S1.33]} & \cite{van2003teaching} & Deepening the kernel language approach for curricula \\
{[S1.34]} & \cite{willcock2004formalization} & Formalization of "concepts" for generic programming \\
\bottomrule
\end{tabular}}
\end{table*}

\begin{table*}[h!]
\centering
\caption{Primary studies addressing RQ2: conceptual primitives and unifying mathematical frameworks}
\label{tab:rq2_studies}
\scalebox{0.9}{
\begin{tabular}{@{}llp{0.6\textwidth}@{}}
\toprule
\textbf{ID} & \textbf{Reference} & \textbf{Primary focus} \\
\midrule
{[S2.1]} & \cite{abramsky1997semantics} & Game semantics for modelling computation \\
{[S2.2]} & \cite{abrial1984programming} & Programming as a formal mathematical exercise \\
{[S2.3]} & \cite{amato2009algebraic} & Category theory (indexed categories) for declarative languages \\
{[S2.4]} & \cite{arbab2005abstract} & Abstract Behaviour Types as a foundation for components \\
{[S2.5]} & \cite{banuatre2000gamma} & The Gamma chemical reaction model \\
{[S2.6]}  & \cite{bauer2015programming} & Algebraic effects and handlers as a programming model \\
{[S2.7]} & \cite{bohm1985automatic} & Synthesis of programs using typed lambda-calculus \\
{[S2.8]} & \cite{bracha1992programming} & Mixins, modularity, and multiple inheritance \\
{[S2.9]} & \cite{buneman1991structural} & Structural recursion for database query languages \\
{[S2.10]} & \cite{cruz2016choreographies} & Minimal primitives for Turing-complete choreographies \\
{[S2.11]} & \cite{cui2005ats} & Combining programming with theorem proving (dependent types) \\
{[S2.12]} & \cite{diaconescu1994category} & Category-based semantics for logic programming \\
{[S2.13]} & \cite{dos2005generic} & Mathematical foundations of generic programming \\
{[S2.14]} & \cite{finkelstein2003new} & Categorical framework for declarative programming \\
{[S2.15]} & \cite{gabbrielli1991modeling} & Modelling answer constraints in constraint logic programs \\
{[S2.16]} & \cite{janson1994akl} & Principles for a multiparadigm language (AKL) \\
{[S2.17]} & \cite{jifeng2000unifying} & UTP with healthiness conditions \\
{[S2.18]} & \cite{lanese2020general} & General approach for deriving reversible semantics \\
{[S2.19]} & \cite{lepper2014foundations} & Unified foundations for functional and data-flow programming \\
{[S2.20]} & \cite{li1994defining} & Unified framework for defining program constructs \\
{[S2.21]} & \cite{lisper2005data} & Formalism for data parallelism in functional programming \\
{[S2.22]} & \cite{maietti2009minimalist} & Minimalist two-level foundation for constructive mathematics \\
{[S2.23]} & \cite{manna2003mathematical} & Mathematical theory of computation for program verification \\
{[S2.24]} & \cite{mccarthy1959basis} & Mathematical theory of computation with recursive functions \\
{[S2.25]} & \cite{meseguer1993logical} & Rewriting Logic for concurrent objects \\
{[S2.26]} & \cite{mycroft1996integration} & Algebraic view on the integration of paradigms \\
{[S2.27]} & \cite{nuka2006mechanising} & Mechanising UTP in a theorem prover \\
{[S2.28]} & \cite{oeyen2022graph} & Graph-based formal semantics for reactive programming \\
{[S2.29]} & \cite{ohori1990representing} & Monads (category theory) for modelling object identity \\
{[S2.30]} & \cite{Oliveira2014} & F-Algebras as a unifying abstraction for FP and OOP \\
{[S2.31]} & \cite{patterson2022semantic} & Semantic models for multi-language type soundness \\
{[S2.32]} & \cite{plotkin2013handling} & Foundational theory of algebraic effects for compositional semantics \\
{[S2.33]} & \cite{plotkin1981structural} & Foundational compositional framework (Structural Operational Semantics) \\
{[S2.34]} & \cite{santos2006object} & Formalization of object-orientation in UTP \\
{[S2.35]} & \cite{saraswat1992higher} & Linear concurrent constraint programming with state change \\
{[S2.36]} & \cite{scholz1995four} & Minimal concurrency primitives in a Process monad \\
{[S2.37]} & \cite{silbermann1992domain} & Domain-theoretic approach to functional/logic programming \\
{[S2.38]} & \cite{sipser1996introduction} & Theory of computation (automata, computability, complexity) \\
{[S2.39]} & \cite{spinellis1994programming} & Modeling paradigms as object classes for multiparadigm structuring \\
{[S2.40]} & \cite{wegner1990concepts} & Foundational concepts of OOP (objects, classes, inheritance) \\
\bottomrule
\end{tabular}}
\end{table*}

\end{document}